\documentclass[a4paper,11pt]{article}
\usepackage{pos}

\title{Science Potential and Technical Design of the IceCube-Gen2 Surface Array}
\ShortTitle{IceCube-Gen2 Surface Array}

\manuallySeparateAuthors
\author[a,b]{Frank G.~Schroeder}
\author[c]{ for the IceCube-Gen2 Collaboration}

\affiliation[a]{Bartol Research Institute, Department of Physics and Astronomy, University of Delaware,\\
  Sharp Lab, 104 The Green, Newark DE, 19716, United States of America}
\affiliation[b]{Institute for Astroparticle Physics, Karlsruhe Institute of Technology (KIT),\\
Postfach 3640, 76021 Karlsruhe, Germany}
\affiliation[c]{Full author list at \href{https://icecube.wisc.edu/collaboration/authors/\#collab=IceCube-Gen2&date=2024-12-15&formatting=web}{https://icecube.wisc.edu/collaboration/authors/}}

\emailAdd{fgs@udel.edu}

\abstract{IceCube-Gen2, the next generation extension of the IceCube Neutrino Observatory at the South Pole, offers a unique scientific potential for cosmic-ray physics at PeV to EeV energies complementing the main science case of neutrino astronomy. The cosmic-ray science case will be enabled by a surface array on top of an extended optical array deep in the polar ice. The optical array measures TeV muons of air showers, and the surface array primarily measures the electromagnetic shower component and low-energy muons. The design of the surface array foresees scintillation panels providing a full-efficiency threshold for near-vertical proton showers of 0.5 PeV and radio antennas increasing the measurement accuracy for the electromagnetic shower component in the energy range of the Galactic-to-extragalactic transition. Compared to IceCube, the aperture for air showers measured in coincidence with the surface and optical arrays will increase by a factor of 30, due to the larger area and angular acceptance in zenith angle. The science potential includes both, the particle physics of air showers, such as prompt muons, and the astrophysics of the highest energy Galactic cosmic-rays, enabled by the higher sensitivity for the mass composition and anisotropy of cosmic rays, and by the search for PeV photons. This proceeding summarizes the science case and design of the surface array as presented in the recently released IceCube-Gen2 Technical Design Report: \href{https://icecube-gen2.wisc.edu/science/publications/tdr/}{https://icecube-gen2.wisc.edu/science/publications/tdr/}}

\FullConference{7th International Symposium on Ultra High Energy Cosmic Rays (UHECR2024)\\
 17-21 November 2024\\
Malargüe, Mendoza, Argentina\\}

\begin{document}
\maketitle

\section{Introduction}
\begin{figure}[b]
\begin{center}
    \includegraphics[width=0.99\linewidth]{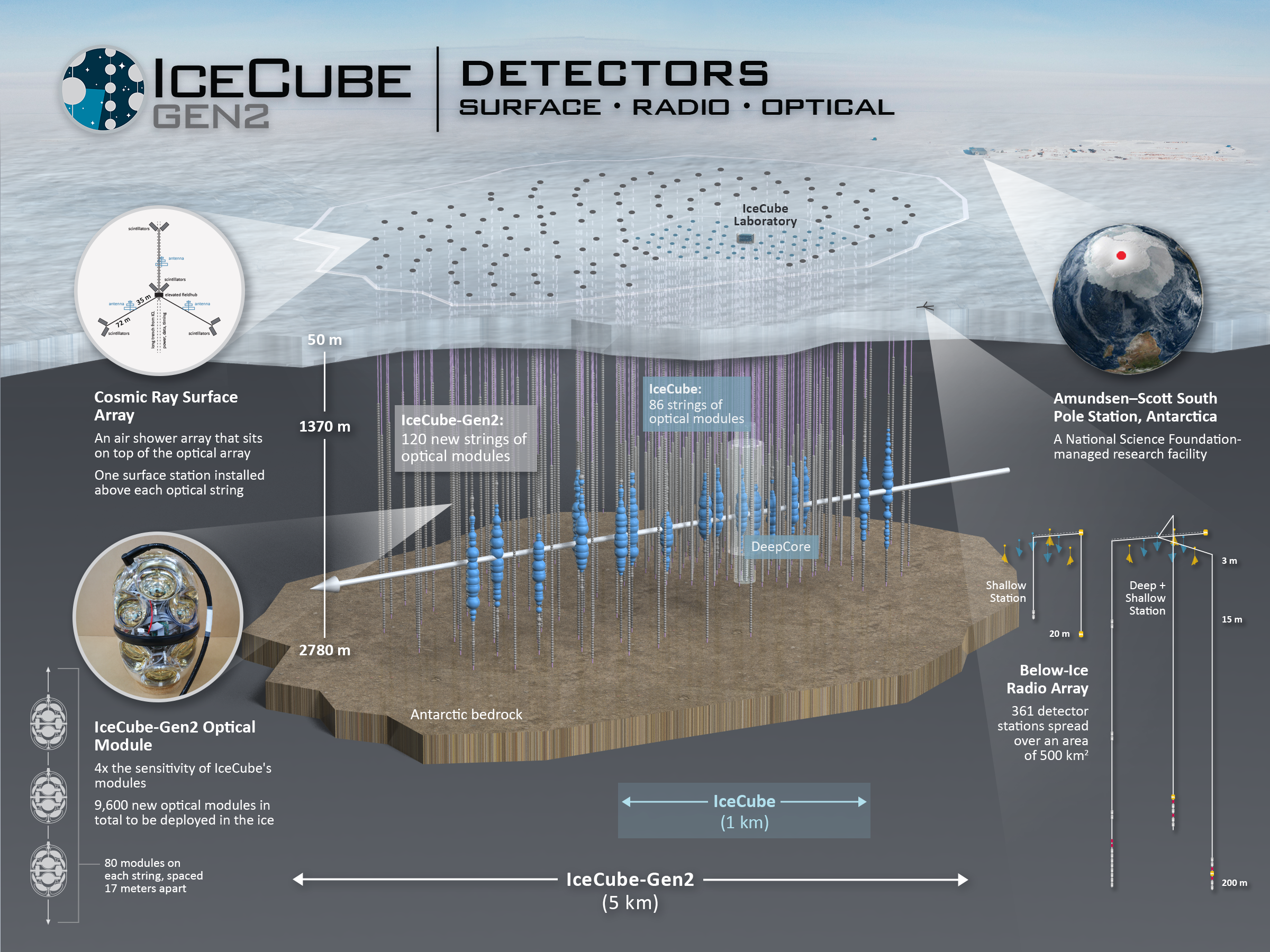}
    \caption{Sketch of IceCube-Gen2 with its Optical Array underneath the Surface Array of elevated radio antennas and scintillation panels, and a Radio Array for ultra-high-energy neutrinos extending beyond the area shown in the figure.}
    \label{fig_IceCubeGen2}
\end{center}
\end{figure}

IceCube-Gen2 will be a next generation multi-messenger detector at the South Pole, extending the IceCube Neutrino Observatory.
It will build on IceCube's successful concept combining a deep optical detector in the Antarctic ice with a surface array for air showers.
The optical array will be comprised of 120 strings of digital optical modules instrumenting an ice volume of about $8\,$km$^3$ with the primary science goal of neutrino astronomy in the TeV to PeV energy range. 
On top of the optical array, a $6\,$km$^2$ surface array comprised of elevated radio antennas and scintillation panels will enhance the neutrino science and enable additional science goals regarding the particle physics and astrophysics of PeV to EeV cosmic rays. 
Additionally, a $500\,$km$^2$ radio array comprised of antennas in the firn will boost the sensitivity to ultra-high-energy neutrinos (see Figure~\ref{fig_IceCubeGen2}).

The recently completed Technical Design Report (TDR)~\cite{IceCubeGen2TDR} complements the previous white paper~\cite{IceCube-Gen2:2020qha} on the science case and provides a detailed description of the plans.
Here, we focus on the science case of IceCube-Gen2 Surface Array which builds on the improved instrumentation together with a 30-fold increase of surface-deep coincident air shower events compared to IceCube.

\begin{figure}[t]
\begin{center}
    \includegraphics[width=0.99\linewidth]{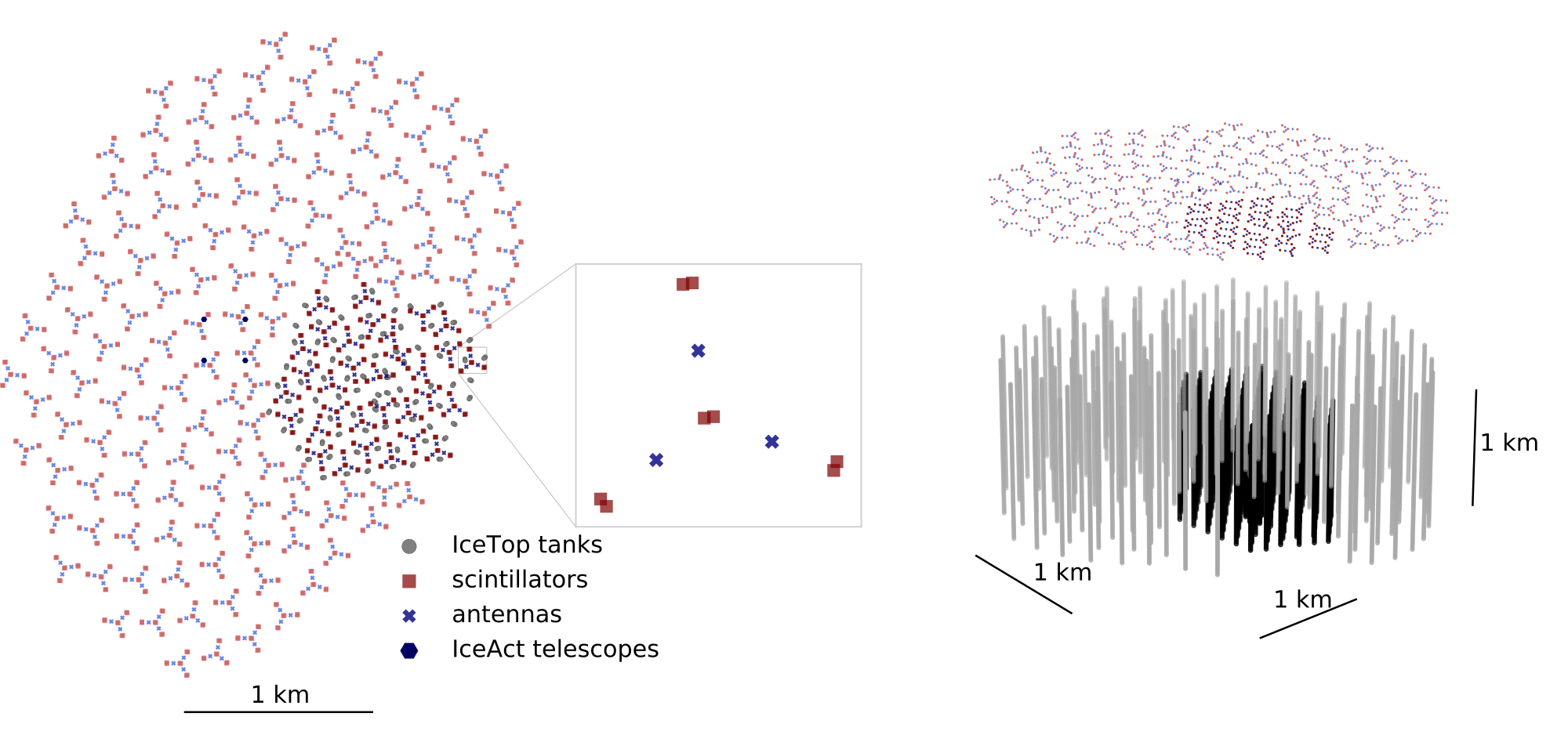}
    \caption{Layout of the IceCube-Gen2 Surface Array: above each of the 120 strings of the optical array, there will be one surface station consisting of three elevated radio antennas and eight scintillation detectors. 
    Additional surface stations enhance the existing IceCube detectors.
    }
    \label{fig_SurfaceLayout}
\end{center}
\end{figure}

\begin{figure}[t]
\begin{center}
    \includegraphics[height=7cm]{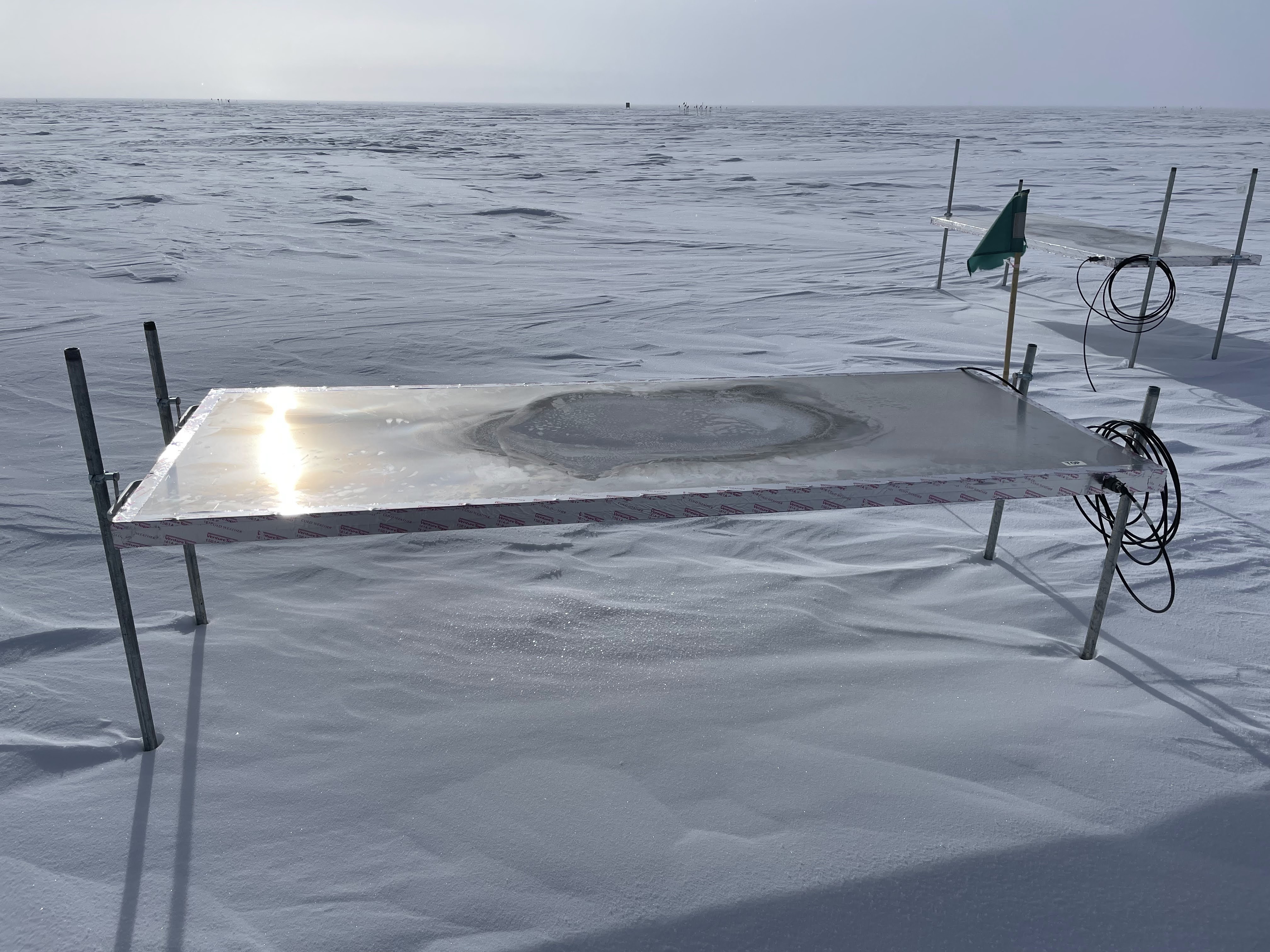}
    \hfill
    \includegraphics[height=7cm]{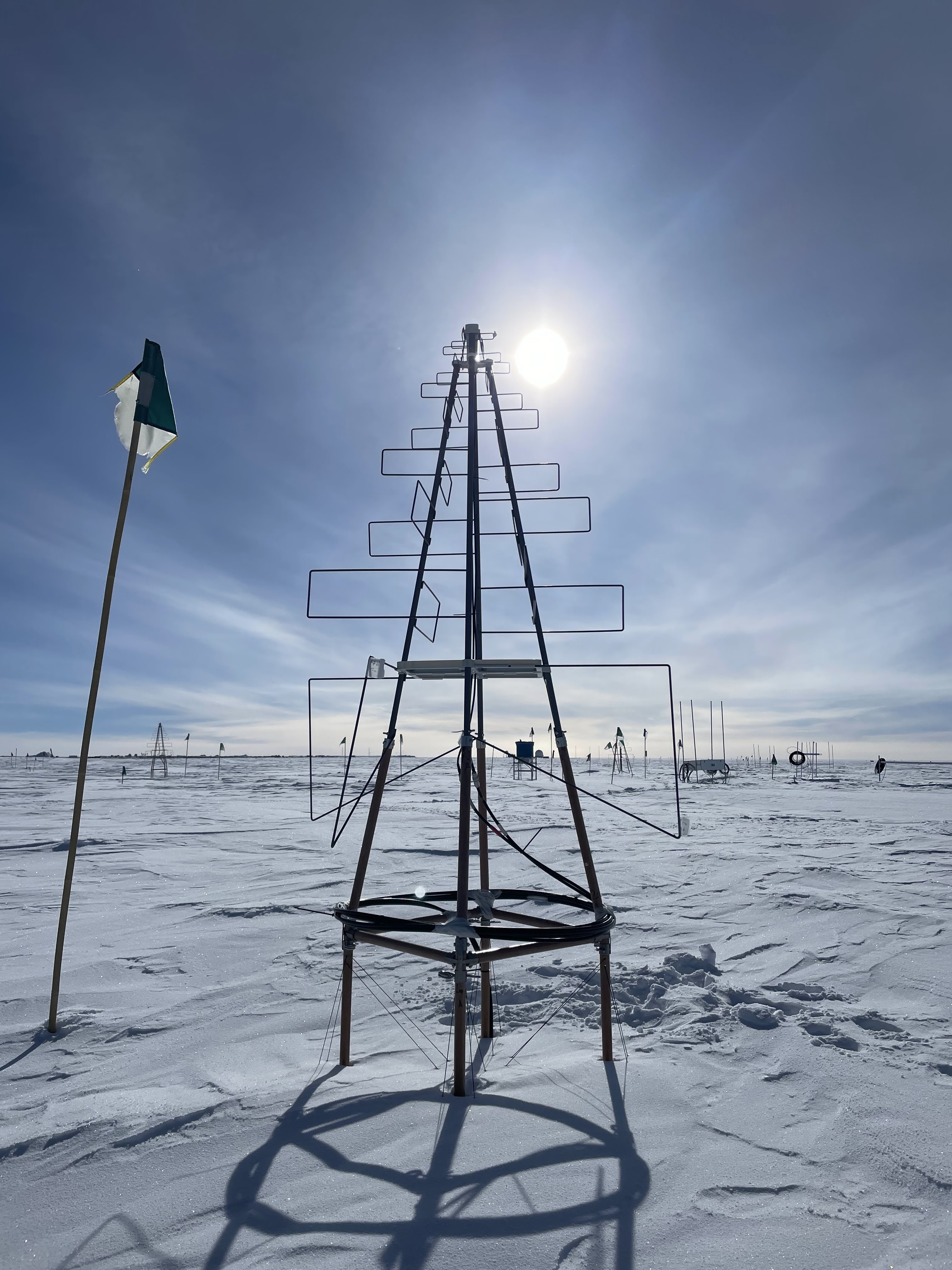}
    \caption{Photos of the detectors of the prototype station that were deployed in January 2020 taken almost three years later in December 2022. Left: One of the four pairs of scintillation panels. Right: One of the three SKALA v2 radio antennas. The concept of elevating the detectors successfully avoids snow accumulation above the detectors and enables to easily raise them every few years corresponding to the rising snow level at the South Pole. Photos courtesy of Roxanne Turcotte.}
    \label{fig_detectors}
\end{center}
\end{figure}

\section{Technical Design of the IceCube-Gen2 Surface Array}
The reference design of the IceCube-Gen2 surface array covers the footprint above the deep optical array with stations of 8 elevated scintillation panels and 3 elevated radio antennas, each, with a prototype station operating successfully at the South Pole for several years~\cite{Shefali:2022kvp} (see Figure~\ref{fig_SurfaceLayout} for the array layout and Figure~\ref{fig_detectors} for photos of the detectors).
The concept of the surface array is inspired by IceTop~\cite{IceCube:2012nn}, the surface array of IceCube comprised of ice-Cherenkov detectors. 
The new detector design brings two major advantages over IceTop: 
First, elevating the detectors on poles avoids snow coverage above the detectors, as snow coverage turned out to increase the detection threshold and systematic uncertainties of IceTop.
Second, the combination of scintillation panels and radio antennas instead of only ice-Cherenkov tanks features a lower detection threshold of $0.5\,$PeV for protons and a higher measurement accuracy in the energy range of the Galactic-to-extragalactic transition from several $10$ PeV to a few EeV. 

The high measurement accuracy is enabled by the radio antennas which are sensitive to the electromagnetic shower component, and provide an accurate measurement of the energy and the depth of the shower maximum, $X_\mathrm{max}$~\cite{Huege:2016veh,Schroder:2016hrv}.
The scintillation panels are a dedicated development for IceCube~\cite{IceCube:2021ydy}, building on the design of similar detectors at AugerPrime~\cite{PierreAuger:2023yab}, but using silicon photomultipliers, taking advantage of the low noise level at the cold ambient temperatures at the South Pole.
For the radio antennas, the SKALA design developed for the Square Kilometer Array~\cite{7297231} fulfills all requirements, in particular coverage of the design band of $70-350\,$MHz, wide zenith coverage, and a low backlobe of the gain pattern, which implies low systematic uncertainties with variations of the snow level.
IceAct telescopes~\cite{IceCube:2019yev} in the center improve the sensitivity to air showers of lower energy.
Furthermore, the layout of the array exploits synergies with the optical array.
In particular, the data-acquisition electronics of a surface station will be located in the same fieldhub as the electronics of the corresponding string of the optical array,  connected to the same power grid and sharing the WhiteRabbit timing and communication system.

With these detectors, the surface array provides a measurement of the air-shower energy, depth of shower maximum ($X_\mathrm{max}$), and its GeV muon content, while the optical array measures the TeV muon content of the air shower and, both together, form a unique detector for cosmic-ray air showers in the PeV to EeV energy range.

\section{Science Case of the IceCube-Gen2 Surface Array}
The Surface Array enhances the IceCube-Gen2 science case in multiple ways. 
In addition to enhancing neutrino astronomy, IceCube-Gen2 will make unique contributions to the particle and astrophysics of Galactic cosmic rays and the yet enigmatic transition to extragalactic cosmic rays.
Finally, IceCube-Gen2 will also be sensitive to gamma rays and feature discovery potential for sources of PeV photons.

\begin{figure}[t]
\begin{center}
\includegraphics[width=0.6\textwidth]{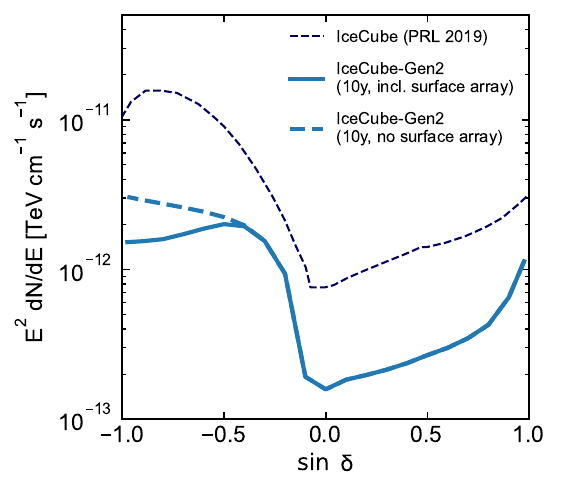}
\end{center}
\caption{Integral sensitivities of IceCube-Gen2~\cite{IceCubeGen2TDR} and IceCube~\cite{IceCube:2019cia} for the discovery of a neutrino point source (for 5$\sigma$ discovery potential after 10 years, assuming a spectrum with a power-law index of $-2$).
IceCube's sensitivity includes the IceTop surface veto, which is visible as an improvement as low declination (dip in the curve to the left-hand side). 
The sensitivity improvement by the surface veto is larger for IceCube-Gen2 due to the wider zenith-angle range of surface-optical coincidences.}
\label{fig_10y_int_sens_surface}
\end{figure}

\subsection{Enhancing Neutrino Astronomy}
The surface array will provide a veto for downgoing muon tracks entering the optical array to discriminate events induced by neutrinos from air showers. 
This enhances the sensitivity to neutrinos particularly for low declinations, enabling the detection of additional PeV neutrino observations from the northern sky (see Figure~\ref{fig_10y_int_sens_surface}). 
The effect of the veto will be even larger than indicated in this figure, as also some air showers uncontained in the surface array will still be detectable and can also be vetoed. 
Although this additional benefit is hard to quantify, IceTop has proven to be effective in purifying the real-time alerts of IceCube in this way, as several downgoing alert candidates were detected by IceTop, indicating a cosmic-ray instead of a neutrino origin~\cite{Amin2021}.
Finally, the surface array will contribute to a better understanding of atmospheric backgrounds (muons and atmospheric neutrinos), as these depend on hadronic interactions and the cosmic-ray flux and mass composition, all of which are targets of IceCube-Gen2's cosmic-ray science.

\begin{figure}[t]
\begin{center}
\includegraphics[width=0.75\textwidth]{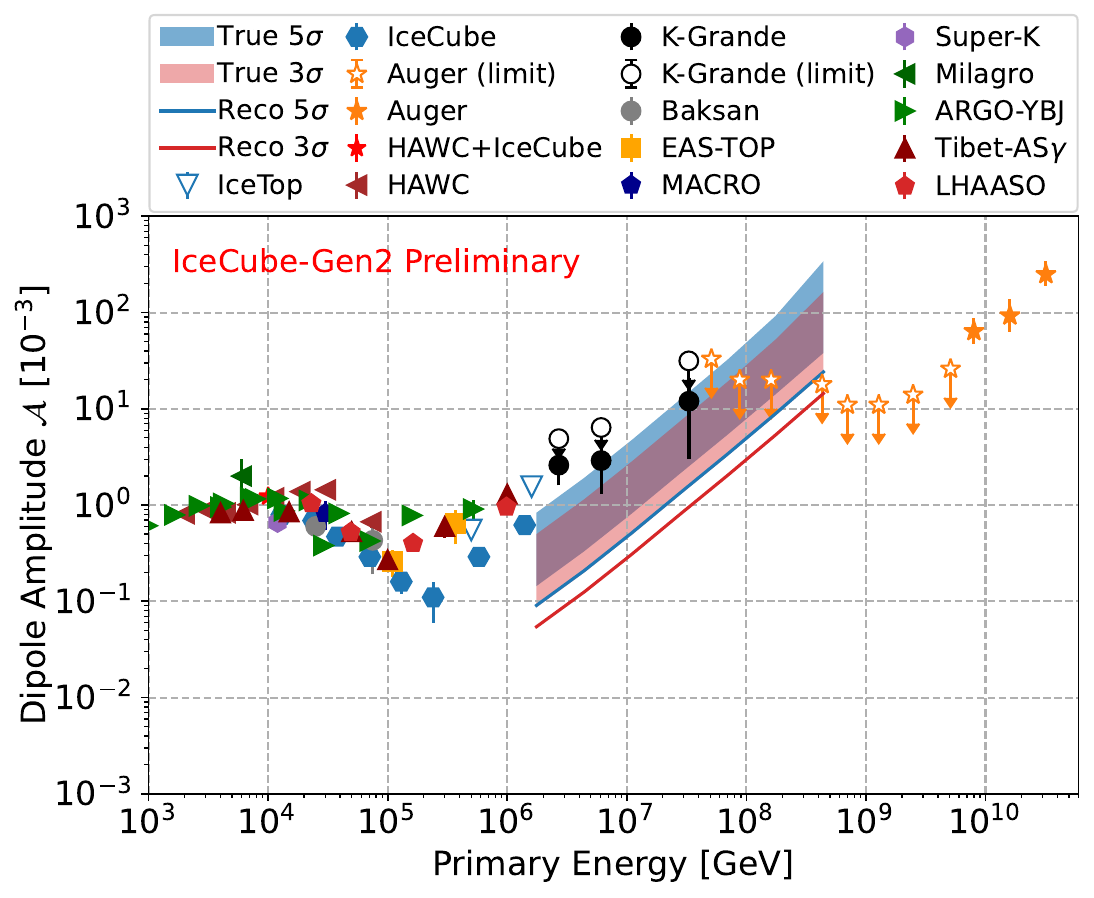}
\end{center}
\caption{Sensitivity of the IceCube-Gen2 Surface Array to the dipole amplitude of the cosmic-ray anisotropy.}
\label{fig_anisotropy}
\end{figure}

\subsection{Cosmic-Ray Astrophysics}
The IceCube-Gen2 surface array will measure cosmic rays of all types with full efficiency from below a PeV to several EeV and provide measurements of the energy spectrum, mass composition, and anisotropy.
In particular, in the energy range of the second knee to the ankle, it will feature unprecedented accuracy thanks to the combination of the radio antennas, scintillation panels, and deep detector providing complementary measurements of the same air showers.
Also, the increase in exposure provides high discovery potential for anisotropies of Galactic cosmic rays at higher energies than discovered so far~\cite{IceCube:2016biq}, reaching up to and possibly beyond the energy of the second knee around $100\,$PeV (see Figure~\ref{fig_anisotropy})~\cite{IceCube-Gen2:2023dxt}.

\subsection{Particle Physics with Air Showers}
As IceCube, also IceCube-Gen2 will be a unique lab to study particle physics in air showers, thanks to the combination of the surface and optical arrays~\cite{Verpoest:2023qmq}.
As a major advantage over IceCube, the $X_\mathrm{max}$ measurements for individual showers by the radio antennas will provide a new dimension to study cosmic-rays, e.g., by  selecting deep showers initiated likely by protons, or by testing whether hadronic interaction models simultaneously reproduce the muon content and $X_\mathrm{max}$ of measured air showers. 
Moreover, the 30 times larger aperture for surface-deep coincidences, will enable a new way to study prompt muon production, as previous studies had statistical limitations~\cite{IceCube:2015wro}.

\begin{figure}[t]
\begin{center}
\includegraphics[width=0.99\textwidth]{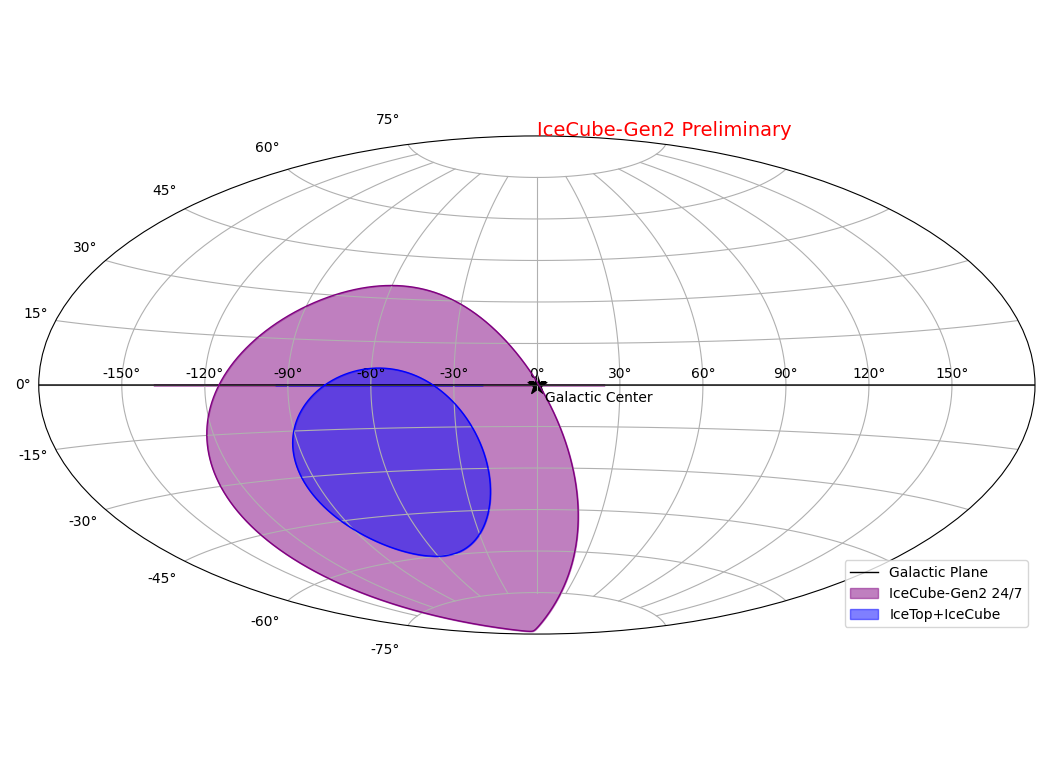}
\end{center}
\caption{Field-of-view in galactic coordinates for air-shower events with the shower axis crossing both, the surface and optical arrays of IceCube / IceCube-Gen2, which provides excellent gamma-hadron separation in the PeV energy range.}
\label{fig_FOV}
\end{figure}

\subsection{PeV Photon Search}
IceCube has shown that the combination of a surface array with a deep array provides excellent gamma-hadron separation because almost all hadronic showers in the PeV energy range feature high-energy muons detectable with the optical array, but most photon-induced showers do not~\cite{IceCube:2019scr}.
As an advantage over IceCube, IceCube-Gen2 will have a larger field of view for gamma rays: due to the larger surface array, the axis of more inclined air showers will intersect the surface and optical arrays, enabling the observation of a larger range of the Galactic plane (see~Figure~\ref{fig_FOV}). 
As the aperture of IceCube-Gen2 for photons is larger than that of LHAASO, which recently detected PeV photons from several sources in other regions of the Galactic Plane~\cite{LHAASO:2021gok}, it is therefore likely that also IceCube-Gen2 will see sources of PeV photons within a few years of operation.
This will provide an additional handle to investigate the sources of the most energetic Galactic cosmic rays.

\section{Conclusion}
IceCube-Gen2 will continue IceCube's success story into the era of neutrino and multimessenger astronomy. 
For both IceCube and IceCube-Gen2, the surface arrays are much more than a veto enhancing neutrino astronomy.
In particular, the combination of the surface and optical arrays enables unique measurements of air showers and a rich science case of cosmic-ray physics up to EeV energies. 
Due to the improved detector design featuring elevated radio antennas in addition to scintillation panels, IceCube-Gen2 will not only increase the statistics by an order of magnitude.
More importantly, it will also provide a different quality of data, due to the combination of radio and muon detection~\cite{Holt:2019fnj}.
In particular, the simultaneous measurement of the air-shower energy and $X_\mathrm{max}$ at the surface and the TeV muons deep in the ice promises unprecedented separation power for the mass groups of primary cosmic rays~\cite{Flaggs:2023exc}.
Thanks to this major increase of both, statistics and accuracy of cosmic-ray measurements, IceCube-Gen2 will thus be a critical experiment to answer open questions regarding the particle and astrophysics of the highest energy Galactic cosmic rays~\cite{Coleman:2022abf}.

\bibliographystyle{ICRC}
\bibliography{bibliography}

\providecommand{\href}[2]{#2}\begingroup\raggedright\begin{thebibliography}{10}

\bibitem{IceCubeGen2TDR}
R.~Abbasi {\em et~al.} {\em IceCube-Gen2 Website} (2024) .
  \url{https://icecube-gen2.wisc.edu/science/publications/tdr/}.

\bibitem{IceCube-Gen2:2020qha}
{\bfseries IceCube-Gen2} Collaboration, M.~G. Aartsen {\em et~al.}
  \href{http://dx.doi.org/10.1088/1361-6471/abbd48}{{\em J. Phys. G} {\bfseries
  48} no.~6, (2021) 060501}.

\bibitem{Shefali:2022kvp}
{\bfseries IceCube} Collaboration, S.~Shefali
  \href{http://dx.doi.org/10.22323/1.398.0055}{{\em PoS} {\bfseries
  EPS-HEP2021} (2022) 055}.

\bibitem{IceCube:2012nn}
{\bfseries IceCube} Collaboration, R.~Abbasi {\em et~al.}
  \href{http://dx.doi.org/10.1016/j.nima.2012.10.067}{{\em Nucl. Instrum. Meth.
  A} {\bfseries 700} (2013) 188--220}.

\bibitem{Huege:2016veh}
T.~Huege \href{http://dx.doi.org/10.1016/j.physrep.2016.02.001}{{\em Phys.
  Rept.} {\bfseries 620} (2016) 1--52}.

\bibitem{Schroder:2016hrv}
F.~G. Schr\"oder \href{http://dx.doi.org/10.1016/j.ppnp.2016.12.002}{{\em Prog.
  Part. Nucl. Phys.} {\bfseries 93} (2017) 1--68}.

\bibitem{IceCube:2021ydy}
{\bfseries IceCube} Collaboration, R.~Abbasi {\em et~al.}
  \href{http://dx.doi.org/10.22323/1.395.0225}{{\em PoS} {\bfseries ICRC2021}
  (2021) 225}.

\bibitem{PierreAuger:2023yab}
{\bfseries Pierre Auger} Collaboration, A.~Abdul~Halim {\em et~al.}
  \href{http://dx.doi.org/10.1088/1748-0221/18/10/P10016}{{\em JINST}
  {\bfseries 18} no.~10, (2023) P10016}.

\bibitem{7297231}
E.~de~Lera~Acedo {\em et~al.},
  \href{http://dx.doi.org/10.1109/ICEAA.2015.7297231}{``Evolution of skala
  (skala-2), the log-periodic array antenna for the ska-low instrument,''} in
  {\em 2015 International Conference on Electromagnetics in Advanced
  Applications (ICEAA)}, pp.~839--843.
\newblock 2015.

\bibitem{IceCube:2019yev}
{\bfseries IceCube} Collaboration, M.~G. Aartsen {\em et~al.}
  \href{http://dx.doi.org/10.1088/1748-0221/15/02/T02002}{{\em JINST}
  {\bfseries 15} no.~02, (2020) T02002}.

\bibitem{IceCube:2019cia}
{\bfseries IceCube} Collaboration, M.~G. Aartsen {\em et~al.}
  \href{http://dx.doi.org/10.1103/PhysRevLett.124.051103}{{\em Phys. Rev.
  Lett.} {\bfseries 124} no.~5, (Feb., 2020) 051103}.

\bibitem{Amin2021}
{\bfseries IceCube} Collaboration, N.~M.~B. Amin
  \href{http://dx.doi.org/10.13140/RG.2.2.20356.24969}{{\em Journal of Physics:
  Conference Series} (2021) }.

\bibitem{IceCube:2016biq}
{\bfseries IceCube} Collaboration, M.~G. Aartsen {\em et~al.}
  \href{http://dx.doi.org/10.3847/0004-637X/826/2/220}{{\em Astrophys. J.}
  {\bfseries 826} no.~2, (2016) 220}.

\bibitem{IceCube-Gen2:2023dxt}
{\bfseries IceCube-Gen2} Collaboration, W.~Hou {\em et~al.}
  \href{http://dx.doi.org/10.22323/1.444.0354}{{\em PoS} {\bfseries ICRC2023}
  (2023) 354}.

\bibitem{Verpoest:2023qmq}
{\bfseries IceCube} Collaboration, S.~Verpoest
  \href{http://dx.doi.org/10.22323/1.444.0207}{{\em PoS} {\bfseries ICRC2023}
  (2023) 207}.

\bibitem{IceCube:2015wro}
{\bfseries IceCube} Collaboration, M.~G. Aartsen {\em et~al.}
  \href{http://dx.doi.org/10.1016/j.astropartphys.2016.01.006}{{\em Astropart.
  Phys.} {\bfseries 78} (2016) 1--27}.

\bibitem{IceCube:2019scr}
{\bfseries IceCube} Collaboration, M.~G. Aartsen {\em et~al.}
  \href{http://dx.doi.org/10.3847/1538-4357/ab6d67}{{\em Astrophys. J.}
  {\bfseries 891} (8, 2019) 9}.

\bibitem{LHAASO:2021gok}
{\bfseries LHAASO} Collaboration, Z.~Cao {\em et~al.}
  \href{http://dx.doi.org/10.1038/s41586-021-03498-z}{{\em Nature} {\bfseries
  594} no.~7861, (2021) 33--36}.

\bibitem{Holt:2019fnj}
E.~M. Holt, F.~G. Schr\"oder, and A.~Haungs
  \href{http://dx.doi.org/10.1140/epjc/s10052-019-6859-4}{{\em Eur. Phys. J. C}
  {\bfseries 79} no.~5, (2019) 371}.

\bibitem{Flaggs:2023exc}
B.~Flaggs, A.~Coleman, and F.~G. Schr\"oder
  \href{http://dx.doi.org/10.1103/PhysRevD.109.042002}{{\em Phys. Rev. D}
  {\bfseries 109} no.~4, (2024) 042002}.

\bibitem{Coleman:2022abf}
A.~Coleman {\em et~al.}
  \href{http://dx.doi.org/10.1016/j.astropartphys.2023.102819}{{\em Astropart.
  Phys.} {\bfseries 149} (2023) 102819}.

\end{thebibliography}\endgroup

\section*{Acknowledgement}

{
The authors gratefully acknowledge the support from several agencies and institutions. In addition to those acknowledged with the \href{https://icecube-gen2.wisc.edu/science/publications/tdr/}{full authorlist}, we specially acknowledge:\\
This project has received funding from the European Research Council (ERC) under the European Union’s Horizon 2020 research and innovation programme (grant agreement No 802729). 
}

\end{document}